\documentstyle[pra,aps,twocolumn]{revtex} 

\newcommand{\da}{^{\dagger}}

\newcommand{\nonpsi}{\tilde{\psi}}
\newcommand{\r}{{\bf r}}

\newcommand{\angles}[1]{\langle#1\rangle}

\newcommand{\partder}[2]{\frac{\partial{#1}}{\partial{#2}}}

\newcommand{\be}{\begin{equation}}
\newcommand{\ee}{\end{equation}}
\newcommand{\ba}{\begin{eqnarray}}
\newcommand{\ea}{\end{eqnarray}}
\newcommand{\bastar}{\begin{eqnarray*}}
\newcommand{\eastar}{\end{eqnarray*}}
\newcommand{\la}{\label}

\newcommand{\taver}[1]{\langle#1\rangle}
\newcommand{\promega}{{\mathbf \Omega}}
\newcommand{\elab}{\angles{E}}

\begin{document}
\draft
\title{On the effect of the thermal gas component to the stability of 
vortices in trapped Bose-Einstein condensates}
\author{S.~M.~M.~Virtanen and M.~M.~Salomaa}
\address{Materials Physics Laboratory, 
Helsinki University of Technology\\
P.~O.~Box 2200 (Technical Physics), FIN-02015 HUT, Finland}
\date{\today}

\maketitle
\begin{abstract}
We study the stability of vortices in trapped single-component Bose-Einstein
condensates within self-consistent mean-field theories---especially we
consider the  Hartree-Fock-Bogoliubov-Popov theory and its recently proposed 
gapless extensions. It is shown that for sufficiently repulsively 
interacting systems the anomalous negative-energy modes related to vortex 
instabilities are
lifted to positive energies due to partial filling of the vortex core
with noncondensed gas. Such a behavior implies that within these theories
the vortex states are eventually stable against transfer of condensate
matter to the anomalous core
modes. This self-stabilization of vortices, shown to occur under 
very general circumstances, is contrasted to the predictions
of the non-self-consistent Bogoliubov approximation, which is known
to exhibit anomalous modes for all vortex configurations and thus implying 
instability of these states. In addition, the shortcomings 
of these approximations in 
describing the properties of vortices are analysed, and the need of
a self-consistent theory taking properly into account the coupled dynamics
of the condensate and the noncondensate atoms is emphasized.
\end{abstract}
\hspace{5mm}

%\narrowtext

\section{Introduction}

The realization of Bose-Einstein condensation in trapped, 
dilute atomic gases \cite{first_exp1,first_exp2,first_exp3}
has opened up unique ways to investigate how weak particle 
interactions affect the quantum-statistical phenomenon of 
bosonic condensation. Especially interesting is the question whether
the interactions can sustain superfluidity in such systems.
A characteristic property of superfluids is the ability to 
support dissipationless flow, a feature which is manifested in 
microscopic length scales as the stability of quantized vortices even in
fluids confined by
nonrotating vessels. This aspect on the superfluid properties of
dilute atomic Bose-Einstein condensates (BECs) has been under 
vigorous theoretical investigation during recent years 
\cite{stringari1,dalf_stringari,dodd_eigenfr,IM_stab0,rokhsar,fetter_lt_stab,svid_fetter_stab,IM_stab1,IM_stab2,IM_stab3,castin_dum,herra_puu,stringari2,butts_rokhsar,feder,garciat,fetter_svid_lett,linn_fetter,svid_fetter_2000,feder2,mcgee_holland,garcia3,our_letter1,our_letter2,garcia4} (for a review, see Ref.\ \cite{fetter_review}),
and has been further motivated by the experimental advances in 
creating and observing vortex structures and their dynamics in these systems
\cite{vortex_exp1,vortex_exp2a,vortex_exp2b,vortex_prec_exp,vortex_exp_ketterle_1,vortex_exp_ketterle_2,hodby}. However, the majority of the analysis
has been carried out
 within the zero-temperature Bogoliubov approximation, which
neglects the effects of the noncondensed gas component always present
in interacting systems. Especially, in addition to providing a dissipative
mechanism for the condensate vortex state \cite{fedichev}, 
the noncondensate component also modifies the excitation spectrum 
of the system, and
may thus in principle affect the stability of vortices.

In discussing the energetic stability,
one has to distinguish between global, thermodynamic stability and
stability against small deformations, or local stability. A physical
system is defined to be 
locally energetically stable if its state is a local minimum
of the free-energy functional under the given external 
constraints---if the state is also a global minimum of energy, it is
thermodynamically stable \cite{lyapunov_stability}. 
In studying the superfluid properties of
dilute atomic BECs, it is crucial to find out whether these systems
support persistent currents without application of any external 
drives---especially, whether vortices are stable in condensates confined
by nonrotating, stationary traps. In this context, the relevant stability
criterion is the local energetic stability, because such 
vortex states are not thermodynamically stable even in 
superfluids---the global energy minimum in a stationary trap 
always corresponds to a nonrotating state. 
On the other hand, local stability of 
bosonic systems is associated with the positivity of the energies of the
normal modes. Our interest hence is to determine the 
local stability, or metastability, of vortex states by investigating their 
normal mode excitation spectra.

The local stability of vortices in harmonically trapped atomic
BECs has been addressed mainly using the zero-temperature Bogoliubov 
scheme, in which the noncondensed gas is
neglected. It has been shown that within the 
Bogoliubov approximation (BA), vortex states in nonrotating
traps always have anomalous excitations with positive norm but negative energy 
(with respect to the condensate state), which 
essentially correspond to a translational displacement of the 
vortex from the center of the condensate and/or bending of the vortex line
\cite{dodd_eigenfr,IM_stab0,rokhsar,garciat,fetter_svid_lett,feder2}.
The negative energies of these excitations imply that vortex states
are energetically unstable: In the presence of any initial perturbation and
dissipative mechanism, these modes would start to be excited through 
the transfer
of matter from the condensate state. In addition to possible bending and
deformation of the vortex, the negative-energy displacement instability
mode implies that the vortex is prone to start spiralling away from the
center of the condensate \cite{rokhsar}. 
By computing the energy of the condensate particles
as a function of the vortex displacement from the condensate center
\cite{mcgee_holland},
it has been verified that it is energetically favourable for the vortex
to eventually spiral out of the condensate and be annihilated.
In conclusion, when the thermal gas component in the system is
neglected, the mean-field theory predicts vortices to be energetically 
unstable.

At temperatures comparable to the condensation temperature $T_{\sc bec}$,
a substantial portion of the gas is noncondensed and the validity of
the instability prediction given by the BA is not obvious.
However,
surprisingly enough, for vortex states its validity is questionable
also in the zero-temperature limit. In interacting systems the noncondensate
density never vanishes exactly, but the residual noncondensate density
in the zero-temperature limit is practically 
negligibly small for weakly interacting
condensates. However, because of the existence of the negative-energy modes
within the BA for vortex states, in the presence of dissipation 
a nonnegligible portion of the condensate
always ``leaks'' to noncondensate populations of these anomalous modes.
The consequent noncondensate density concentrated in the vortex core region, 
in turn, may in principle affect
the local stability of the vortex state by changing the anomalous mode
energies themselves. A similar phenomenon is observed for
two-component condensate systems, in which an irrotational condensate
component can stabilize vortex states \cite{mcgee_holland}. The crucial
property determining whether such a stabilization is also possible
due to finite noncondensate density, is the amount of noncondensate
required to lift the anomalous modes to positive energies: If the
lifting effect is too weak, the whole condensate vortex state
will leak to macroscopic occupations of the anomalous modes, and
the vortex will be destroyed. 

Previously, the self-stabilization mechanism described above has been
demonstrated in the case of an infinitely long, axisymmetric, static
vortex line located in the center of a harmonically trapped condensate.
It has been shown that within the self-consistent Popov approximation (PA) 
\cite{griffin_form}
and its recently proposed gapless extensions G1 and G2
\cite{gener1,gener2}, the noncondensate 
in the vortex core lifts the anomalous modes to positive energies for 
thermally equilibrated vortex states even in the zero-temperature limit
\cite{IM_stab2,our_letter1}. However, even slightly
off-axis vortices are not stationary, but rather move about the condensate
typically at velocities, for which the assumption of adiabaticity
implicitly made when applying stationary self-consistent mean-field 
theories to describe time-dependent phenomena is highly questionable
\cite{our_letter2}. This raises the question of applicability of the
results obtained for axisymmetric vortex states to the general off-axis vortex
case. In this paper we generalize the previous results for the PA, G1 and
G2 by showing that
the anomalous modes of a circularly precessing off-axis vortex line
penetrating a sufficiently strongly interacting, trapped 
condensate are lifted to positive energies as the noncondensate partially
fills the vortex core region---the noncondensate fraction
required for such stabilization depends on the effective strength
of the repulsive interactions in the system, but it is of the order of
the ratio of the vortex core volume to the volume of the whole condensate.

In conclusion, the Bogoliubov approximation predicts vortices in harmonically
trapped BECs to be energetically unstable, while the self-consistent
Popov, G1 and G2 mean-field approximations (and other approximations 
having a similar
structure; we refer to them with the term Popov-like approximations (PLAs)) 
predict them to be stable. This discrepancy can be contrasted
to the experimental observations of vortex precession in spherical 
condensates \cite{vortex_prec_exp}. Theoretically, it has been shown that the 
precession direction of a vortex is related to the energy of the vortex 
displacement mode: If the
mode is anomalous with a negative energy, the vortex precesses in the
direction of the condensate flow, and vice versa 
\cite{fetter_svid_lett,linn_fetter,svid_fetter_2000}. Except for a minority
of so-called rogue vortices, the precession direction observed supports the
predictions of the BA rather than those of the self-consistent approximations.
As the results derived in this paper suggest that the
non-adiabaticity in the time-development of moving vortices is not 
responsible for the discrepancy between the experimental
observations and the predictions of the PLAs, we are 
essentially left with two possibilities: Either dissipation
in the experimental conditions has been
too weak for the vortex states to be sufficiently
properly thermalized, or these self-consistent approximations fail in 
describing the low-energy collective modes of vortex configurations.
These possibilities are discussed in more detail in 
Sec.\ \ref{Discussion and Conclusions} where we discuss the 
shortcomings of the above-mentioned models in describing 
vortex states, and stress
the need for an analysis within a self-consistent formalism that takes into 
account the dynamics of the condensate and the noncondensate on an 
equal footing.

The structure of this paper is the following. In Sec.\ \ref{General Formalism}
the general formalism of time-dependent Hartree-Fock-Bogoliubov-Popov
equations in a rotating frame of reference is presented. In 
Sec.\ \ref{Precessing Vortex Line} this formalism is applied to the case
of a condensate containing a precessing off-axis vortex line, and a
lower bound for the quasiparticle energies is derived. Furthermore,
by applying this lower bound to the anomalous modes localized in the
vortex core region, we show that such modes are lifted to positive
energies as the noncondensate density in the core exceeds a certain
value. Finally, the implications and proper interpretation of these
results are discussed in Sec.\ \ref{Discussion and Conclusions}.

\section{General Formalism}
\label{General Formalism}

We model a dilute bosonic gas with the usual effective grand-canonical
Hamiltonian, which takes the second-quantized form
\ba
\la{hamiltonian}
\hat{H}&=&\int d\r\:\psi\da(\r,t){\mathcal H}_{\bf \Omega}(\r)\psi(\r,t)
\nonumber\\
&+&\frac{g}{2}\int d\r\:\psi\da(\r,t)\psi\da(\r,t)\psi(\r,t)\psi(\r,t)
\ea
in a frame rotating with the angular velocity ${\bf \Omega}$ w.r.t.\ 
an inertial coordinate system. Above, $\psi(\r,t)$ denotes the boson
field operator, and
\be
\la{ham_one}
{\mathcal H}_{\bf \Omega}(\r)\equiv
-\frac{\hbar^2}{2m}\nabla^2+V_{\rm tr}(\r)-\mu-{\bf \Omega}\cdot {\bf L}
\ee
is the one-particle Hamiltonian, with $V_{\rm tr}(\r)$ denoting the 
external trapping potential, $\mu$ the chemical potential, and 
${\bf L}$ the angular momentum operator. The interaction between the 
particles is modelled by the effective contact potential
$V_{\rm int}(\r-\r')=g\delta(\r-\r')$, with the coupling constant
$g$ related to the vacuum $s$-wave scattering length $a$ 
through $g=4\pi\hbar^2a/m$.
The dynamics of the system is determined by the Heisenberg
equation of motion
\be
\la{heisenberg}
i\hbar\partder{}{t}\psi(\r,t)
={\mathcal H}_{\bf \Omega}(\r)\psi(\r,t)+g\psi\da(\r,t)\psi(\r,t)\psi(\r,t).
\ee
Following the spontaneous symmetry breaking approach, we use the 
Bogoliubov decomposition
\be
\psi(\r,t)=\Phi(\r,t)+\nonpsi(\r,t)
\ee
to split the field operator to the sum of the $c$-number condensate 
wave function $\Phi(\r,t)=\taver{\psi(\r,t)}$ (with $\taver{\cdots}$ we
signify the time-dependent nonequilibrium average)
and the noncondensate,
i.e., excitation field operator $\nonpsi(\r,t)$.
The expectation value of Eq.\ (\ref{heisenberg}) yields, when terms
containing the cubic noncondensate operator product or the anomalous 
average $\taver{\psi\psi}$ are neglected according to the Popov
mean-field scheme, the generalized Gross-Pitaevskii (GP) equation
\be
\la{tgp}
i\hbar\partder{}{t}\Phi(\r,t)={\mathcal L}_{\bf \Omega}(\r,t)\Phi(\r,t)
-gn_{\rm c}(\r,t)\Phi(\r,t)
\ee
for the condensate wave function. Above,
${\mathcal L}_{\bf \Omega}(\r,t)\equiv 
{\mathcal H}_{\bf \Omega}(\r)+2gn(\r,t)$, and
the condensate, noncondensate and total gas densities are, respectively,
denoted as
\begin{mathletters}
\la{densities}
\ba
n_{\rm c}(\r,t)&=&|\Phi(\r,t)|^2,\\
\la{noncond}
\tilde{n}(\r,t)&=&\taver{\nonpsi\da(\r,t)\nonpsi(\r,t)},\\
\la{total_dens}
n(\r,t)&=&n_{\rm c}(\r,t)+\tilde{n}(\r,t);
\ea
\end{mathletters}
the chemical potential $\mu$ is implicitly determined by the
condition
\be
N=\int d\r\:n(\r,t),
\ee
where $N$ is the total number of particles in the system.
Furthermore, Eqs.\ (\ref{heisenberg})--(\ref{densities}) imply in
the Popov scheme the equation of motion
\be
\la{noncondheis}
i\hbar\partder{}{t}\nonpsi(\r,t)={\mathcal L}_{\bf \Omega}(\r,t)\nonpsi(\r,t)
+g\Phi^2(\r,t)\nonpsi\da(\r,t)
\ee
for the noncondensate operator. This equation can be diagonalized
by expressing the field operator in terms of bosonic quasiparticle
operators $\alpha_n$ and $\alpha\da_n$. The Bogoliubov 
transformation
\be
\la{bogo}
\nonpsi(\r,t)=\sum_n[u_n(\r,t)\alpha_n-v_n^*(\r,t)\alpha\da_n]
\ee
converts the operator equation (\ref{noncondheis}) to the
time-dependent Hartree-Fock-Bogoliubov-Popov (HFBP) equations
\begin{mathletters}
\la{thfb}
\ba
i\hbar\partder{}{t}u_n(\r,t)&=&{\mathcal L}_{\bf \Omega}(\r,t)u_n(\r,t)
-g\Phi^2(\r,t)v_n(\r,t)\\
-i\hbar\partder{}{t}v_n(\r,t)&=&{\mathcal L}_{-{\bf \Omega}}(\r,t)v_n(\r,t)
-g{\Phi^*}^2(\r,t)u_n(\r,t)
\ea 
\end{mathletters}
for the quasiparticle amplitudes $u_n(\r,t)$ and $v_n(\r,t)$. In addition,
for the Bogoliubov transformation to be canonical, we must impose the
condition
\be
\la{norm}
\int d\r\, [u_n^*(\r,t)u_{n'}(\r,t)-v_n^*(\r,t)v_{n'}(\r,t)]=\delta_{nn'}
\ee
for the amplitudes---this time-independent normalization can 
straightforwardly be shown to
be consistent with Eqs.\ (\ref{thfb}). Furthermore, Eqs.\ (\ref{noncond})
and (\ref{bogo}) imply for the noncondensate density the self-consistency 
relation 
\ba
\la{self_rho}
\tilde{n}&=&\sum_{nn'}[f_{nn'}(u_n^*u_{n'}+
v_n^*v_{n'})-2\text{Re}\{g_{nn'}u_n v_{n'}\}\nonumber\\
&&\qquad+\delta_{nn'}|v_n|^2],
\ea
where
%\begin{mathletters}
\be
f_{nn'}(t)=\taver{\alpha\da_n\alpha_{n'}}\,,\qquad
g_{nn'}(t)=\taver{\alpha_n\alpha_{n'}}
\ee
%\end{mathletters}
denote the normal and anomalous quasiparticle distribution functions;
here and henceforth we often do not explicitly denote the arguments
of functions, unless they are needed for clarity.

Neglect of the anomalous average in the Popov scheme is a rather crude
method to obtain a theory with gapless spectrum in the homogeneous limit,
as required by Goldstone's theorem. Effects of the anomalous average
can be taken into account by renormalizing the two-body interaction
potential to a $T$-matrix including many-body effects. Effectively,
this can be done by replacing the coupling constant $g$ by suitable spatially
varying coupling functions in the mean-field 
equations \cite{gener1,gener2}.
In the following analytic treatment we assume, for simplicity, the effective
interaction to be constant, and only finally briefly 
discuss the more general situation.

\section{Precessing Vortex Line}
\label{Precessing Vortex Line}

Consider a repulsively interacting 
boson condensate containing a precessing vortex
line, trapped by a static, axisymmetric potential. 
For simplicity, we assume the vortex to move in a circular orbit about
the trap symmetry axis, with a constant angular velocity $\promega$
directed along the trap axis. Such a precession motion has been observed
in the experiments, and it is in agreement with vortex dynamics given
by Eq.\ (\ref{tgp}) when the noncondensate density is neglected 
\cite{prec_orbit}. In addition, we assume the system to be thermalized 
in the sense that the quasiparticle distribution functions $f_{nn'}$ and
$g_{nn'}$ are constant in time. In such a case, the mean-field
Hamiltonian is time-independent in the frame rotating with the
angular velocity $\promega$ w.r.t.\ the laboratory
coordinate system. Consequently, in such a frame Eqs.\ (\ref{tgp}) and 
(\ref{thfb}) reduce to the stationary equations
\be
\la{gp}
{\mathcal L}_{\bf \Omega}(\r)\Phi(\r)
-gn_{\rm c}(\r)\Phi(\r)=0
\ee
for the condensate wave function and
\begin{mathletters}
\la{hfb}
\ba
\la{uhfb}
{\mathcal L}_{\promega}(\r)u_n(\r)
-g\Phi^2(\r)v_n(\r)&=&E'_n u_n(\r)\\
\la{vhfb}
{\mathcal L}_{-\promega}(\r)v_n(\r)
-g{\Phi^*}^2(\r)u_n(\r)&=&-E'_n v_n(\r)
\ea 
\end{mathletters}
for the quasiparticle eigenenergies $E'_n$ and amplitudes 
$u_n(\r)$, $v_n(\r)$. Although stationary, these equations are not to 
be confused with the adiabatic approximation
for systems with slow time dependence \cite{our_letter2}. The adiabatic 
approximation fails if the quasiparticle states do not 
follow the moving vortex core rigidly, but the eigenequations 
(\ref{hfb}) describe the time dependence of the system in principle exactly.
The essential difference between these formalisms is the terms 
$\promega\cdot\mathbf{L}$ contained
in Eqs.\ (\ref{hfb}), but absent in the equations corresponding to the
adiabatic approximation---it is these
terms that account for the deformation of the vortex structure due
to its motion \cite{our_letter2}.

Although considerably less cumbersome than the general time-dependent HFBP
formalism, the stationary Eqs.\ (\ref{hfb}) have not been formulated nor
solved so far for
self-consistent solutions describing off-axis vortices. However, in the
following we show that the essential question concerning the local energetic
stability of such vortex states can be studied analytically under very
general circumstances. It is to be noted that the thermalization and 
the energetic stability of the system are 
determined by its excitation spectrum in
the nonrotating, i.e., laboratory frame (LF) of reference, since we
assume that the gas is confined with a static external potential and that
the thermal component as a whole is nonrotating \cite{baym}. 
Hence, in order to investigate the issue of local stability,
we have to study the quasiparticle energies in the LF, instead of the
rotating frame eigenenergies $E'_q$. 
Strictly speaking, as the mean-field Hamiltonian 
in the LF has explicit time dependence via the mean fields,
it does not possess stationary solutions, and, consequently,
a well-defined spectrum---in the laboratory frame only the
\emph{expectation values}
of the quasiparticle energies are well-defined. For the same reason,
such a system can not attain exact thermodynamic equilibrium. 
However, if the thermalization rate of the gas substantially exceeds the 
kinetic rates related to the motion of the 
vortex, the system is thermalized in the sense that its dynamics becomes
adiabatic to a high degree of accuracy \cite{adiabaticity}, 
and the quasiparticle distribution functions approach time-independent 
values.

\subsection{Lower Bound for Quasiparticle Energies}

In order to investigate the positivity of the energies of the 
lowest quasiparticle excitations,
we first derive a lower bound for the spectrum of a trapped BEC
within the Popov approximation.

Let $(\Phi, u, v)$ be a self-consistent solution of
Eqs.\ (\ref{gp}) and (\ref{hfb}), corresponding to an eigenenergy
$E'$ and a given set of time-independent quasiparticle 
distribution functions $f_{nn'}$ and $g_{nn'}$.
Eqs.\ (\ref{hamiltonian}) and (\ref{ham_one}) imply that
the expectation value $\elab$ of the quasiparticle energy in the 
nonrotating laboratory frame is related to the 
eigenenergy $E'$ in the rotating frame by the relation 
$\elab=E'+\promega\cdot\angles{{\mathbf l}}$, where 
$\angles{{\mathbf l}}$ denotes the expectation value of the
angular momentum of the quasiparticle state. By expanding the
expectation value 
$\int d\r\:\angles{\nonpsi\da{\mathbf L}\nonpsi}$
of the angular momentum related to the noncondensate 
in terms of the quasiparticle amplitudes and distribution functions,
one finds that the angular momentum of the system is changed by the amount
$\int d\r\:(u^*{\mathbf L}u+v{\mathbf L}v^*)$ as the corresponding
quasiparticle occupation number is increased by one. Thus, we find
\be
\la{e_ep}
\elab=E'+\int d\r\,[u^*(\promega\cdot{\mathbf L})u
+v(\promega\cdot{\mathbf L})v^*].
\ee
On the other hand, integration of the sum of Eq.\ (\ref{uhfb})
multiplied by $u^*(\r)$ and the complex conjugate of Eq.\ (\ref{vhfb})
multiplied by $v(\r)$ yields 
\ba
\la{euv}
\int d\r\, [u^*({\mathcal L}_0-\promega\cdot{\mathbf L})u
+v({\mathcal L}_0-\promega\cdot{\mathbf L})v^* - 2 g\Phi^2u^*v]\nonumber\\
=E'\int d\r\,(|u|^2-|v|^2)=E',
\ea
where we have used the relation 
${\mathcal L}_{\promega}={\mathcal L}_0-\promega\cdot{\mathbf L}$,
the fact that ${\mathbf L}^*=(\r\times\hbar\nabla/i)^*=
-\r\times\hbar\nabla/i=-{\mathbf L}$, and
the normalization condition (\ref{norm}) for the quasiparticle amplitudes.
Now, combination of Eqs.\ (\ref{e_ep}) and (\ref{euv}) yields the
relation
\be
\la{e1}
\elab=\int d\r\, (u^*{\mathcal L}_0 u+v{\mathcal L}_0 v^*-
2g\Phi^2u^*v)
\ee
for the quasiparticle energy in the laboratory frame. It is to be noted
that this expression has only an implicit dependence on the vortex precession
velocity $\promega$, via the fact that the amplitudes $u(\r)$ and $v(\r)$ 
satisfy Eqs.\ (\ref{hfb}). 

In order to derive a simple lower-bound expression for the quasiparticle
energy, we first note that $\int d\r\:\Phi^2u^*v$ is real valued---this can be
inferred from Eq.\ (\ref{e1}), where all the other terms are
real valued \cite{complex_eig}. Thus we have
\ba
\int d\r\:\Phi^2u^*v
%\leq\bigg|\int d\r\:\Phi^2u^*v\bigg|
&\leq& \int d\r\:|\Phi|^2|u||v|\nonumber\\
&\leq& \frac{1}{2}\int d\r\:n_c(|u|^2+|v|^2),
\ea
since $(|u|-|v|)^2=|u|^2+|v|^2-2|u||v|\geq 0$. Combination of this upper
bound with Eq.\ (\ref{e1}) yields, due to the positivity of the coupling
constant $g$, the inequality
\begin{eqnarray}
\label{ein1}
\elab&\geq&\int d\r\,[u^*{\mathcal L}_0 u + v{\mathcal L}_0 v^*
-(|u|^2+|v|^2)g n_c]\nonumber\\
&=&\int d\r\,[u^*(-\hbar^2\nabla^2/2m)u
+v(-\hbar^2\nabla^2/2m)v^*]\nonumber\\
&+&\int d\r\,(|u|^2+|v|^2)(V_{\rm tr} + g n_c + 2 g \tilde{n} - \mu).
\end{eqnarray}
The kinetic energy terms may be discarded from this lower bound due
to their positivity, and we find
\be
\la{ein2}
\elab> \int d\r\,\rho\,(V_{\rm tr} + g n_c + 2 g \tilde{n} - \mu),
\ee
where $\rho(\r)=|u(\r)|^2+|v(\r)|^2$ denotes the density distribution of
the quasiparticle state. Using the Thomas-Fermi (TF) approximation
\be
\la{tf}
n_c^{\rm (TF)} = \frac{1}{g}(\mu-V_{\rm tr}-2g\tilde{n})
\ee
for the condensate density in the strong-interaction limit, 
we can write the second factor in the
integral of the lower bound (\ref{ein2}) as
\be
\la{sec_factor}
V_{\rm tr} + g n_c + 2 g \tilde{n} - \mu
= g (n_c-n_c^{\rm (TF)}).
\ee
Since we are mainly interested in
the anomalous quasiparticle states which are localized in the vicinity 
of the vortex core, it is useful to split the spatial integral in the
inequality (\ref{ein2}) to contributions from the core region
$C$ and its complement $C'$. To be specific, we define the vortex center
line to consist of points in which $\Phi(\r)=0$, and the core region $C$ to
contain points whose distance from the center line is less than the
healing length $\xi=(8\pi n_0 a)^{-1/2}$, where $n_0$ denotes the local total 
density of a corresponding irrotational 
system. Especially, in the strong interaction Thomas-Fermi limit we have
\be
\la{n_0_tf}
n_0\simeq\frac{1}{g}(\mu-V_{\rm tr}),
\ee 
provided that the temperature is not so close to $T_{\sc bec}$ that the 
thermal gas component would dominate over the condensate.
Utilizing the positivity of $\rho$, $n_c$ and $g$,
we find Eqs.\ (\ref{ein2}) and (\ref{sec_factor}) to imply 
\ba
\la{ein3}
\elab &>& 
\int_C d\r\, \rho\,(V_{\rm tr}+2g\tilde{n}-\mu)\nonumber\\
&&-g\max_{C'}\{n_c^{\rm (TF)}-n_c\} \int_{C'} d\r\, \rho,
\ea
where $\max\{\cdots\}_{C'}$
denotes the maximum value in the region $C'$. Furthermore,
by defining the quantity
\be
w=\frac{\int_C d\r\, \rho}{\int_{C'} d\r\, \rho}
\ee
describing the degree of localization of the quasiparticle state $(u,v)$
into the vortex core region, we can write the lower bound (\ref{ein3})
for the quasiparticle energy in the more compact form
\be
\la{ein4}
\elab > \int_C d\r\, \rho\left[V_{\rm tr}+2g\tilde{n}-\mu
-g\max_{C'}\{n_c^{\rm (TF)}-n_c\}/w\right].
\ee
It is to be noted that this lower bound holds for all quasiparticle states,
when the corresponding density distributions $\rho$ and localization
degrees $w$ are used in the integrand.

\subsection{Anomalous Modes}

Suppose that the condensate containing a vortex line has an
anomalous quasiparticle mode (AM) with a negative energy in 
the laboratory frame. In such a case the system is energetically unstable, 
because it can, in the presence of
dissipative mechanisms, lower its free energy by exciting the AM
with matter originating from the condensate. It has been argued
that such ``drain'' of matter into the AM would lead to
the formation of a binary condensate from the vortex core, and a 
simultaneous spiralling of the vortex out of the condensate 
\cite{rokhsar}. However, it is to be noted that when the noncondensate 
mean field is taken into account in a self-consistent 
manner, the quasiparticle energies depend
on the distribution functions $f_{nn'}$ and $g_{nn'}$. Especially,
excitation of an AM increases the corresponding
occupation number, and thus changes the energy of the mode itself.
If the change in the energy is large enough, the anomalous mode may
attain positive energies, in which case the ``collapse'' of the condensate
would cease. Consequently, the vortex would be stabilized by the
self-interaction of the noncondensate in the core. 

More specifically, since the quasiparticle density distribution $\rho$ is
positive, Eq.\ (\ref{ein4}) implies that a given mode can
be a negative-energy AM only if the expression in the brackets is
negative somewhere in the core region, i.e.,
\be
\la{AM_in_1}
\tilde{n} < \frac{1}{2g}\left(\mu-V_{\rm tr}
+g\max_{C'}\{n_c^{\rm (TF)}-n_c\}/w\right)
\ee
somewhere in $C$.

In order to estimate the vortex core localization degrees $w$ of the
anomalous modes, we note that the quasiparticle motion is 
essentially determined by the potential
\be
V_{\rm eff}=V_{\rm tr} + 2g(n_c+\tilde{n})-\mu
\ee
---this is the effective Hartree potential appearing in 
Eqs.\ (\ref{thfb}) and (\ref{hfb}). For noncondensate density 
distributions not too far from the form corresponding to a thermalized
state, $V_{\rm eff}$ is negative in the vortex core region,
and positive elsewhere \cite{rokhsar,IM_stab2}. Consequently, for the energy
of the mode to be negative, its density has to be
concentrated in the core region, thus yielding a localization
degree $w\gtrsim 1$. This heuristic reasoning for the localization
degree is supported by explicit numerical computations
for axisymmetric vortex states of axially harmonically trapped
condensates, resulting in values $w\approx 0.7$
in the large-$N$ limit, and  $w\approx 1.0$ for systems
in the noninteracting limit \cite{computations}. It is to be noted that
for vortex precession velocities that substantially exceed the 
adiabaticity limit for rigid vortex core motion \cite{our_letter2}, 
the quasiparticles may be deformed and the localization degree
can be smaller than the estimates given above. However,
for the anomalous modes it always has to be of order unity.

For harmonically trapped gases, the effective
strength of the interactions relative to the kinetic energy in the system is 
characterized by the dimensionless quantity $Na/a_{\rm ho}$, where
$a_{\rm ho}=(\hbar/m\bar{\omega})^{1/2}$ is the harmonic oscillator length,
and $\bar{\omega}=(\omega_x\omega_y\omega_z)^{1/3}$ denotes the geometric
mean of the harmonic oscillator frequencies. The Thomas-Fermi approximation
for the condensate density is accurate in the interaction-dominated
regime $Na/a_{\rm ho}\gg 1$. However, as the corrections to the 
chemical potential given by the TF value $\mu_{\rm TF}=gn_{\rm peak}$,
where $n_{\rm peak}$ denotes the maximum density of the gas, are
of the small fractional order 
$(15Na/a_{\rm ho})^{-4/5}\ln (15Na/a_{\rm ho})^{1/5}$ 
\cite{sinha,corr_to_tf},
its accuracy is in general on the order of $10$\% when
$Na/a_{\rm ho}\gtrsim 1$. Thus, in this parameter 
regime we may use the estimate (\ref{n_0_tf}) in the condition (\ref{AM_in_1}).
Furthermore, as the localization degree $w$ of the anomalous mode
is of order unity, the contribution given by the last term in
the inequality (\ref{AM_in_1}) is small, and we get an approximate
condition
\be
\la{AM_in_final}
\tilde{n}(\r) \lesssim \frac{1}{2}n_0(\r)
\ee
for some $\r\in C$, 
when $Na/a_{\rm ho}\gtrsim 1$ and the temperature is not too close
to $T_{\sc bec}$. In other words, since $n_0(\r)$ is approximately the
local maximum density in the vicinity of the vortex, 
the anomalous modes are lifted to positive energies as the
noncondensate in the core fills half of the core volume.

This result, which applies also to unharmonically trapped condensates,
shows that within
the PA even partial filling of the vortex core with non-condensate
is sufficient to ``suppress'' the anomalous modes by lifting their
energies to positive values. In fact the approximate inequality
(\ref{AM_in_final}) is rather pessimistic in the
sense that, according to exact numerical solutions,
even much lower noncondensate core densities suffice to
lift the anomalous modes to positive energies \cite{our_letter1}. The analysis 
can also in principle be straightforwardly 
generalized to cover the G1 and G2 approximations,
if and when the effective spatially dependent coupling $g$
is not substantially reduced in the vortex core region.
Returning to  the question of vortex stability, these results show that 
the ``collapse''
of the condensate to the anomalous core modes ceases as the noncondensate
density due to these modes partially fills the vortex core and the energies
of the instability modes rise above the energy of the condensate state.
In the presence of strong enough
dissipation, this self-stabilization is a direct
consequence of thermalization. All in all, we find that Popov-like
self-consistent approximations predict vortices to become locally 
energetically stable in the course of thermalization.

\section{Discussion and Conclusions}
\label{Discussion and Conclusions}

The Bogoliubov approximation consisting of the mean-field Gross-Pitaevskii
equation and the Bogoliubov normal-mode equations has been 
widely used to describe the
properties of vortices and, especially, their stability in dilute atomic BECs.
Predictions derived from it for the vortex precession direction and 
frequency, together with the critical trap rotation frequencies for 
vortex nucleation, agree well with experimental results. However, the
BA neglects the effects of the noncondensate.
At temperatures $T\gtrsim T_{\sc bec}/2$ a considerable fraction of the
gas is noncondensed and the BA is insufficient to model the system. Moreover,
we argue that for vortex states its accuracy is questionable even
at lower temperatures. This is due to the tendency of the condensate
matter to be transferred to noncondensate occupations of the anomalous
negative-energy core modes. Consequently, as the system is thermalized,
the vortex core becomes occupied by a substantial noncondensate density
even in the zero-temperature limit.

The Popov-like approximations, on the other hand, are finite-temperature
mean-field theories that self-consistently take into account the 
thermal gas component. Their predictions for the excitation frequencies of 
irrotational condensates are in excellent agreement with experimental
results at temperatures
$T\lesssim T_{\sc bec}/2$, but at higher temperatures
discrepancies become obvious, as the noncondensate
density becomes more dominant. This failure of the PLAs near $T_{\sc bec}$
has been attributed to the fact that they do not properly treat
the \emph{dynamics} of the noncondensed gas.

On the basis of explicit numerical solutions for axisymmetric vortex states
\cite{IM_stab2,our_letter1}
and the analysis presented in this paper (see also Ref.\ 
\cite{fetter_review}, p.\ R177), we conclude that PLAs imply vortices in
sufficiently properly thermalized and repulsively interacting condensates
to be locally energetically stable in the sense that the excitation spectrum of
the system is positive definite. Especially, these approximations
imply that the vortex precession mode has a positive energy, which in turn
seems to imply that off-axis vortices precess in the direction opposite to
the vortex flow. However, this prediction contradicts experimental
results, which rather show that the majority of vortices precesses in
the direction of the condensate flow. In the light of the success of the
mean-field approximations in describing the physics of 
dilute condensates, the failure
of the mean-field theory itself is an unlikely reason underlying
this discrepancy. Thus we are essentially left with two alternatives:
Either the vortex states observed in the experiments have not been properly
thermalized, or the PLAs drastically fail in describing the lowest
collective modes of the vortex states. 
In fact the observations of vortex precession have so far been 
conducted under conditions in which dissipation in the gas is very
weak, and the degree of thermalization probably low 
\cite{vortex_prec_exp,fedichev}.
It is to be noted that, due to improper thermalization,
the anomalous mode occupations may be negligible and the vortex core
almost void of noncondensate---in such a case the Bogoliubov approximation
should naturally yield more accurate results than the PLAs.
 
In any case, there is also a reason to expect PLAs to fail
in predicting the lowest quasiparticle modes for vortex states even
in the zero-temperature limit. As the vortex state is thermalized,
the noncondensate concentration in the vortex core region becomes 
nonnegligible and, as we have seen, it substantially modifies 
the energies of the lowest quasiparticle states which are 
essentially localized 
in the core. However, when describing the collective modes of the system,
the PLAs do not correctly take into account 
the \emph{dynamics} of this
noncondensate, but treat the thermal gas peak in the core as a static
pinning potential for the vortex. The stabilizing effect of such an
``external'' potential is naturally stronger than that of a
realistic, dynamical noncondensate peak, and the lifting effect of the 
anomalous modes is consequently exaggerated within the PLAs. 
For the same reason, the prediction for the precession direction given
by the PLAs is compromised, because the lowest mode rather describes
precession of the vortex about a static noncondensate peak
than the combined motion of the condensate and the noncondensate
(see also Ref.\ \cite{fetter_review}). Such a failure
of PLAs may be somewhat surprising, as one in general would expect
these approximations to be unreliable only near $T_{\sc bec}$,
where the noncondensate fraction is substantial together with
the resonant contributions to the self-energies \cite{fedi_gora}. 
When vortex states
are considered, however, the dynamics of the 
thermal gas should be taken into account at 
all temperatures, due to the filling of the vortex core with noncondensate
in the course of thermalization.

Consequently, the effects of the thermal gas concentrated in the vortex
core are not properly modelled within the BA nor the PLAs. In order to
reliably find out the structure, spectrum and stability of quantized 
vortices in dilute BECs one has to use a model which treats the dynamics
of the condensate and the thermal gas on an equal footing. Such fully
dynamic approximations for inhomogeneous gases have been developed 
in recent years \cite{fedi_gora,zaremba,morgan_new,second_ord_lett}, 
but they have not yet been applied to study vortex states. 
Also due to the advances in experimental techniques and measurement accuracies
there is a need for investigating vortex states beyond the Bogoliubov level.
Because of the strong inhomogeneities, vortices are potentially
quite efficient testbenches for thermal field theories in modelling
atomic BECs.

In conclusion, we have studied the stability of precessing off-axis vortices
in dilute, trapped Bose-Einstein condensates within the self-consistent
Popov approximation and its extensions G1 and G2. By deriving a lower
bound for the excitation energies, it has been shown that within these
approximations the vortex states are locally energetically stabilized as
the vortex core is partially filled with thermal gas. Furthermore,
as such filling of the vortex core with noncondensate is an inherent
consequence of thermalization and the existence of anomalous modes
for vortex states in pure condensates, we argue that the effects
of the thermal gas for the physics of vortices may not be neglected
even in the zero-temperature limit. Furthermore, the shortcomings 
of the Bogoliubov
approximation and the Popov-like self-consistent approximations in modelling
vortex states have been pointed out and analysed. Essentially, the BA
may be too ``pessimistic'' in predicting vortices to be definitely unstable,
and the PLAs too ``optimistic'' in predicting them to be self-stabilized by
the thermal gas component. We argue that a reliable analysis of the 
stability of quantized vortices in dilute BECs requires a fully dynamic 
formalism---such an investigation remains a challenge for further studies.

\begin{acknowledgements}
We thank T.~P.~Simula for discussions. The Academy of Finland and 
the Graduate School in Technical Physics are appreciated for support.
\end{acknowledgements}


\begin{references}
\bibitem{lyapunov_stability}
Another relevant stability concept is the dynamical or Lyapunov
stability, which encompasses the idea that small perturbations
lead only to small deviations in the time development of the 
state of the system. However, 
energetic stability (both local and global) is a sufficient 
condition for dynamical stability, and we do not consider the latter
separately in this paper.
\bibitem{prec_orbit}
See Refs.\ \cite{feder2,mcgee_holland,vortex_prec_exp}. Our main objective
in extending the stability analysis to the case of moving vortices is
to investigate the effect of non-adiabaticity due to finite velocity of
the vortex line. For this purpose it is sufficient to restrict to consider
only circular vortex motion.
\bibitem{adiabaticity}
Note that here adiabaticity refers to its thermodynamic meaning, 
in contrast to the adiabaticity of the nonlinear quantum dynamics
referred to earlier in this paper.
\bibitem{complex_eig}
Quasiparticle states which obey the normalization condition 
(\ref{norm}) can be shown to have real energies. We do not consider
the complex frequency states occurring for multiquantum vortices due
to their dissociation instability \cite{herra_puu}.
\bibitem{computations}
For the numerical methods used in computing the wavefunctions
and energies of the anomalous modes in axially trapped condensates, see
Ref.\ \protect\cite{our_letter1}.
\end{references}
\end{document}